\documentclass[11pt,a4paper]{article}
\pdfoutput=1

\usepackage{xargs}
\usepackage[pdftex,dvipsnames]{xcolor}
\usepackage[utf8]{inputenc}
\usepackage{uniinput}
\usepackage[a4paper,left=1.1in,right=1.1in,top=1.1in,bottom=1.3in,footskip=0.3in]{geometry}
\usepackage{microtype,booktabs,graphicx,subcaption,enumerate,epsfig,multirow}
\usepackage{amsfonts,amsmath,amssymb,mathtools,commath,mathrsfs}
\usepackage{dsfont,pifont,bbm,units,cancel,slashed,csquotes,tocloft,cite}
\allowdisplaybreaks
\usepackage[colorlinks = true,
linkcolor = blue,
urlcolor  = blue,
citecolor = blue,
anchorcolor = blue]{hyperref}
\hypersetup{pageanchor=false}

\usepackage{scrpage2}
\clearscrheadfoot
\cefoot[--  \pagemark  --]{\pagemark}
\cofoot[--  \pagemark  --]{\pagemark}

\newcommand{\D}{\mathrm{d}}

\begin{document}
	\pagestyle{plain}
	\begin{titlepage}
		\begin{flushright}
			UUITP-46/17\\
		\end{flushright}	
		\bigskip
		\begin{center}
			{\LARGE \bfseries 
				Observational signatures from horizonless black shells imitating rotating black holes \par}
			\bigskip
			{\large  Ulf H. Danielsson and Suvendu Giri}\let\thefootnote\relax\footnote{{\tt 
					ulf.danielsson@physics.uu.se, suvendu.giri@physics.uu.se}}\\
			\bigskip \bigskip
			{\it Institutionen för fysik och astronomi, Uppsala University, \\
				Box 803, SE-751 08 Uppsala, Sweden \\[2mm]
			}
			\bigskip
		\end{center}
		
		\bigskip
		\begin{center}
			{ \bfseries ABSTRACT}\\[3ex]
			\begin{minipage}{0.85\textwidth}
				\small
				In 
				\href{https://arxiv.org/abs/1705.10172}{arXiv:1705.10172}
				it was proposed that string theory replaces Schwarzschild black holes with horizonless thin shells with an AdS interior. In this paper we extend the analysis to slowly rotating black holes, solving the Israel-Lanczos-Sen junction conditions for a rotating shell composed of stringy matter to determine the metric. Outside of the shell we find a vacuum solution that differs from Kerr with a $32\%$ larger quadrupole moment. 
				We discuss the observational consequences and explore the possibility to distinguish between a black shell and a black hole. Promising methods include imaging of the black hole at the center of the Milky Way using the Event Horizon Telescope, precision measurements of stars in close orbits around the central black hole, and future observations of colliding super massive black holes using the space based gravitational wave observatory LISA.
			\end{minipage}
		\end{center}
		\vfill
	\end{titlepage}
	
	\tableofcontents
  \bigskip
  
  \section{Introduction}\label{sec:introduction}
  
  One of the most important problems in black hole physics is the reconciliation of the thermal nature of Hawking radiation with unitary evolution in quantum mechanics. The existence of this \enquote{information paradox} \cite{Hawking:1976ra} relies on black holes having a horizon.
  An early attempt to resolve the paradox was the idea of \emph{black hole complementarity} \cite{tHooft:1991uqr,Susskind:1993if}, which suggests that an infalling and an outgoing observer see complimentary pictures of the information falling into the black hole. In this way there is no loss of information. 
  However, it was suggested that the idea of black hole complementarity might be incomplete \cite{Mathur:2009hf} and the idea of \emph{firewalls} was proposed \cite{Almheiri:2012rt}.
  In parallel, another way to resolve the paradox was proposed \cite{Lunin:2002qf,Giusto:2004id,Mathur:2003hj,Mathur:2004sv} 
  suggesting that a black hole is a collection of \emph{microstates}, each of which describes a smooth horizonless geometry. 
  The conventional picture of a black hole being a central singularity surrounded by empty space, and shielded by a horizon, is thus replaced by an effective description of the statistical ensemble of smooth horizonless geometries (see \cite{Mathur:2005zp} for a review).  There has been extensive work in constructing smooth supergravity solutions corresponding to these \emph{fuzzball} microstates \cite{Bena:2004de,Bena:2012zi,Bena:2017fvm} (see e.g. \cite{Bena:2007kg} for a review).
  
  Other compact horizonless objects that could mimic black holes, and thus resolve the information paradox, have also been proposed (see \cite{Cardoso:2017njb} for a summary of these objects).
  One such possibility that was recently suggested in \cite{Danielsson:2017riq} is that string theory naturally prevents the 
  formation of a horizon at the end point of gravitational collapse, and in this way removes the paradox. Instead of obtaining a black hole, it was argued that  
  a bubble of AdS-space is formed that is contained within a thin shell of branes supporting some stringy matter: a {\it black shell}.
  The tension of these branes, as well as the negative cosmological constant in their interior, is set by the scale of high energy physics making the total energy of the shell and the vacuum inside enormous for an astrophysical black hole. Still, they balance in such a way that the tension of the brane compensates for the negative cosmological constant, and the effective mass measured from outside is just the mass of a traditional black hole.
  Our proposal seems to be different from the fuzzball idea but it would be interesting to explore possible connections.
  
  The black shells we introduced in \cite{Danielsson:2017riq} are similar to the gravastars discussed in \cite{Visser:2003ge}. The crucial difference is that we, motivated by string theory, consider an AdS interior with a negative cosmological constant rather than a de Sitter interior with a positive cosmological constant. In our paper we solved the Israel-Lanczos-Sen junction conditions \cite{Israel1966,doi:10.1002/andp.19243791403,doi:10.1002/andp.19243780505} in detail and found, quite remarkably, that given some basic assumptions about the equations of state of the string matter, the radius of the shell is uniquely determined and turns out to be  the \emph{Buchdahl radius} ($=9R_s/8=9M/4$). Buchdahl \cite{Buchdahl:1959zz} showed that this is the smallest possible radius of a star (modeled as a sphere of incompressible fluid) provided that the pressure is isotropic, and that the density does not increase outward. Schwarzschild \cite{Schwarzschild:1916ae} had derived the same limit when considering the interior metric of an incompressible fluid sphere and Buchdahl generalized the result to any matter distribution with the above properties.
  
  In \cite{Danielsson:2017riq} we also discussed the construction of black shells from string theory in some detail. Starting with an explicit supersymmetric construction of an extremal Reissner-Nordström black hole in Type IIA string theory compactified on $\mathbb{T}^6/\left(\mathbb{Z}_2\times\mathbb{Z}_2\right)$, we argued that these shells could be made of branes with world volume directions both along the shell as well as internal space, together with lower dimensional branes wrapping only the compact internal space. From the point of view of space time, these look like point particles (D0 branes) dissolved in higher dimensional branes, with a gas of open strings stretching between them.
  
  The radius of the shell at equilibrium can be obtained by solving the junction conditions.
  However, to understand the perturbative stability of the shell around such a critical point, one needs to consider quantum effects.
  In \cite{Danielsson:2017riq} we suggested that the shell is heated to the local 
  Unruh temperature since it is accelerating relative to the local inertial frame. This is analogous to how one can argue for the Hawking temperature using an imaginary surface 
  positioned just outside the horizon. One calculates the local Unruh temperature on that surface, which diverges in the limit where the surface approaches the horizon, and then calculates the asymptotic Hawking temperature using the gravitational redshift. Our argument differs in that the shell is now a physical object instead of being just an imaginary surface. Furthermore, it is positioned at a macroscopic distance away from where the horizon would have been. The resulting asymptotic temperature turns out to be a little less than the Hawking temperature.
  
  These shells indeed appear \enquote{black} as they should in order to be compatible with observations. As was shown in \cite{Danielsson:2017riq}, they carry an enormously large number of degrees of freedom (just as in the case of a conventional black hole) but have an extremely low temperature.  By the second law of thermodynamics, the high entropy ensures that infalling matter sticks to the shell, and becomes part of the degrees of freedom of the string gas sitting on top of the shell, while the low temperature ensures that nothing is radiated out for a long time. This ensures that the shells appear black.
  
  In \cite{Danielsson:2017riq} we also proposed a mechanism for the formation of the black shells, and how it guarantees that horizons can never appear. We made use of a background Minkowski space time that was metastable against the nucleation of vacuum bubbles of AdS. The life time of the vacuum was assumed to be extremely long, many times longer than the age of the universe. We argued that if a bubble nucleates on top of a collapsing shell of matter, there is a huge increase in the available phase space, and the low probability of its nucleation is compensated by a similar increase in its entropy. This comes about since the infalling matter has much less entropy compared to the one carried by the gas of strings on the brane. In this way the bubble \enquote{catches} the infalling matter, is trapped and prevented from expanding much further, and finally settles down to a radius dictated by the junction conditions and temperature induced stability.
  
  Most astrophysical black holes are believed to be rotating, often with close to maximal spin \cite{Elvis:2001bn,Wang:2006bz}. In order to test the idea that black holes really are black shells, we therefore need to extend the construction in \cite{Danielsson:2017riq} to rotating black holes. This is technically much more challenging, and in the present work we focus on the limit of slow rotation.
	Here we do not provide an explicit construction of such slowly rotating black shells from string theory but rather use the components that arise from a string construction of black shells \cite{Danielsson:2017riq} (namely the energy-momentum tensor being made of three pieces -- tension ($p=-ρ$), massless gas ($p=ρ/2$) and stiff matter ($p=ρ$)) to construct these shells. We hope to return to an explicit construction from string theory in a future work.
	
  There is substantial literature discussing the space time outside rotating compact objects \cite{DeLaCruz:1968zz,MankoNovikov:1992aa,Abramowicz:2003rc,CastejonAmenedo:1990zz,Hartle:1968si,Uchikata:2015yma}. In the absence of spin (and charge), assuming spherical symmetry, the exterior geometry of such objects is uniquely described by the Schwarzschild metric. 
  In the presence of spin the geometry is unique, provided there is a horizon, and is given by the Kerr solution \cite{Carter:1971zc,Hawking:1971vc}. 
  In our case, the external geometry is cut off at the shell (which sits well outside the horizon at $9M/4$ for the Schwarzschild case), the no-hair theorem does not apply, and there is no reason to expect the metric to be described by the Kerr solution. In fact, the geometry outside axially symmetric rotating objects is not unique, and contributions with non-vanishing Weyl curvature can be added using the Kerr solution as a starting point. 
  Under the assumption of a metastable Minkowski vacuum together with stringy matter, we suggest that the bubble nucleation mechanism of \cite{Danielsson:2017riq} prevents the formation of a Kerr black hole. The geometry outside such rotating black shells differs from Kerr and the stringy matter on the shell uniquely fixes the deviation away from the Kerr geometry.
  Of particular interest is that our construction predicts that the quadrupole moment is about $32\%$ greater than that of the Kerr geometry.
  
  We begin our discussion by presenting the space time geometry inside and outside the shell in section \ref{sec:settingupmetric}. In section \ref{sec:firstorder}, we solve the junction conditions at first order in the rotation parameter $a$ and calculate the stress energy tensor on the shell. We show how this can be understood as a high tension brane with some stringy matter on top. In section \ref{sec:secondorder}, we do the computation at second order in $a$ and obtain the main results of this work. We discuss astrophysical implications of this proposal and suggest ways to test it in section \ref{sec:astrophysical}. In section \ref{sec:conclusions} we end with a summary of our results and an outlook.
  
  \section{Spacetime geometry inside and outside the shell}\label{sec:settingupmetric}
  
  \subsection{Outside the shell}
  The Kerr metric in Boyer-Lindquist coordinates can be written as
  \begin{equation}
	  d s² = -g_{tt} d t² + g_{rr} d r² + g_{ϑϑ} d ϑ² + g_{φφ} d φ² + g_{tφ} d t d φ + g_{φt} d φ d t,
  \end{equation}
  where up to $\mathcal{O}\left(a²\right)$, the metric components are given  by
  \begin{equation}\label{eq:kerr_unperturbed}
  \begin{split}
	  g_{tt} &= 1-\frac{2M}{r}+\frac{2a²M\cos ²ϑ}{r³},\\
	  g_{rr} &= \left(1-\frac{2M}{r}\right)^{-1} 
	  +\frac{a²}{r²}\left[\cos ²ϑ \left(1-\frac{2M}{r}\right)^{-1}-\left(1-\frac{2M}{r}\right)^{-2}\right],\\
	  g_{ϑϑ} &= r² + a² \cos ² ϑ,\\
	  g_{φφ} &= r² \sin ²ϑ + \frac{a²\sin ²ϑ}{r}\left(r+2M\sin ²ϑ\right),\\
	  g_{tφ} &= g_{φt} = \frac{2aM\sin ²ϑ}{r}.
  \end{split}
  \end{equation}
  To generalize beyond the Kerr metric, we will first consider an axially symmetric spacetime of the form
  \begin{equation}\label{eq:ouransatz}
	  d s² = -g_{tt} e^{λ} d t² + g_{rr} e^{ν} d r² + g_{ϑϑ} e^{ν} d ϑ² + g_{φφ} e^{-λ} d φ² + 2 g_{tφ}  d t d φ,
  \end{equation}
  where $g_{μν}$ are the unperturbed quantities from \eqref{eq:kerr_unperturbed}. For a stationary axisymmetric solution, $λ$ and $ν$ can only be functions of $r$ and $ϑ$. Demanding that this metric is a vacuum solution to Einstein's equations, one obtains the following equations of motion up to $\mathcal{O}\left(a²\right)$
  \begin{equation}\label{eq:ourequations}
	  \begin{split}
	  \left(r-3M\right)λ_{,ϑ} + \left(r-M\right)ν_{,ϑ} + r\left(r-2M\right)\cot ϑ\left(λ_{,r}+ν_{,r}\right) &\overset{!}{=} 0,\\
	  \cot ϑ \left(λ_{,ϑ}+ν_{,ϑ}\right)-\left(r-3M\right)λ_{,r}-\left(r-M\right)ν_{,r} &\overset{!}{=}0,\\
	  \left(λ_{,ϑϑ}+ν_{,ϑϑ}\right)+\left(r+M\right)λ_{,r}+\left(r-M\right)ν_{,r}+r\left(r-2M\right)\left(λ_{,rr}+ν_{,rr}\right)
	  &\overset{!}{=}0,\\
	  \left(λ_{,ϑϑ}-ν_{,ϑϑ}\right)+4\left(r-2M\right)λ_{,r}+\cot ϑ\left(λ_{,ϑ}-ν_{,ϑ}\right)
	  +r\left(r-2M\right)\left(λ_{,rr}-ν_{,rr}\right)&\overset{!}{=}0,
	  \end{split}
  \end{equation}
  where subscripts denote partial derivatives, i.e. $λ_{,r}\coloneqq \partial λ/\partial r$ etc.
  
  We demand that the perturbations $λ$ and $ν$ die off as $r→∞$ in order to ensure that the geometry is asymptotically flat. Imposing these boundary conditions, we can pick the following solution to Einstein's equations
  \begin{equation}\label{eq:oursols}
	  \begin{split}
	  λ &= \frac{1}{M²}\bigg[-2\left(2M²-6Mr+3r²\right)\mathscr{T} \cos 2ϑ +2M\left(M-r\right)\left(1+3\cos 2ϑ\right)\\
	  &\qquad\qquad+\left(2M²+2Mr-r²\right)\mathscr{L}\bigg],\\
	  ν &=  \frac{1}{M²}\bigg[2M\left(3M+r+\left(3r-7M\right)\cos 2ϑ\right)\\
	  &\qquad\qquad+\left(-6M²+2Mr+r²+\left(6M²-10Mr+3r²\right)\cos 2ϑ\right)\mathscr{L}\bigg],
	  \end{split}
  \end{equation}
  where
  \begin{equation}\label{eq:arctanh}
	  \mathscr{T}\coloneqq \mathrm{arctanh}\left(\frac{M}{M-r}\right),\qquad \qquad
	  \mathscr{L} \coloneqq \log\left(1-\frac{2M}{r}\right).
  \end{equation}
  The ansatz in \eqref{eq:ouransatz} is of the same form as the Novikov-Manko metric \cite{MankoNovikov:1992aa}, and their solution indeed solves \eqref{eq:ourequations}. It can be seen from \eqref{eq:oursols} that our solution is different from theirs e.g. unlike ours, the Novikov-Manko solution does not have logarithms. 
  The asymptotic behavior is the same at leading order but differs at next to leading order.
  
  Next, we consider another stationary axially symmetric geometry of the form
  \begin{equation}\label{eq:HTansatz}
	  d s² = -g_{tt} e^{χ} d t² + g_{rr} e^{-χ} d r² + g_{ϑϑ} e^{ψ} d ϑ² + g_{φφ} e^{ψ} d φ² + 2 g_{tφ}  d t d φ,
  \end{equation}
  where as before $g_{μν}$ are the unperturbed quantities from \eqref{eq:kerr_unperturbed}, while $χ$ and $ψ$ are functions of $r$ and $ϑ$. Einstein's equations in vacuum are given by
  \begin{equation}
	  \begin{split}
	  2Mχ_{,ϑ}+r\left(r-2M\right)\left(χ_{,rϑ}+ψ_{,rϑ}\right) &\overset{!}{=} 0,\\
	  2rχ_{,r}-2\left(r-M\right)ψ_{,r}+r\left(r-2M\right)\left(χ_{,rr}+ψ_{,rr}\right) &\overset{!}{=}0,\\
	  2\left(χ+ψ\right)+\cot ϑ\left(χ_{,ϑ}+ψ_{,ϑ}\right)+\left(χ_{,ϑϑ}+ψ_{,ϑϑ}\right)+2\left(r-2M\right)χ_{,r}+2\left(r-M\right)ψ_{,r}
	  &\overset{!}{=}0,\\
	  \cot ϑ χ_{,ϑ} + χ_{,ϑϑ} -\left(r-2M\right) \left(2ψ_{,r}+rψ_{,rr}\right) &\overset{!}{=}0,
	  \end{split}
  \end{equation}
  Demanding that the metric is asymptotically flat at large $r$, i.e. $χ$ and $ψ$ go to zero as $r→∞$, we can pick the following solution
  \begin{equation}\label{eq:HTsols}
	  \begin{split}
	  χ &= -\frac{5}{8 M²}\frac{\left(3\cos ²ϑ-1\right)}{r\left(r-2M\right)}
	  \left[M\left(r-M\right)\left(2M²+6Mr-3r²\right)-3r²\left(r-2M\right)²\mathscr{T}\right],\\
	  ψ &= \frac{5}{8 M²}
	  \frac{\left(3\cos ²ϑ-1\right)}{r}\left[M\left(2M²-3Mr-3r²\right)+r\left(6M²-3r²\right)\mathscr{T}\right].
	  \end{split}
  \end{equation}
  where $\mathscr{T}$ is defined in \eqref{eq:arctanh}. The metric thus obtained turns out to be the well known Hartle-Thorne metric \cite{Hartle:1968si}. 
  
  Since Einstein's equations linearize at the order we are working at, we can superimpose the above perturbations to write a combined metric
  \begin{equation}\label{eq:fullexternal}
	  \begin{split}
	  d s²_{\textrm{combined}} =& -g_{tt} e^{ a² \left(q λ+ p χ\right)} d t² + g_{rr} e^{a² \left(q ν- p χ\right)} d r² + g_{ϑϑ} e^{a² \left(q ν+p ψ\right)} d ϑ² \\
	  &+ g_{φφ} e^{a² \left(-q λ+p ψ\right)} d φ² + 2 g_{tφ}  d t d φ,
	  \end{split}
  \end{equation}
  To understand this metric, we can compute its
  \emph{Geroch-Hansen multipole moments} \cite{Geroch:1970cd,Hansen:1974zz}. These are coordinate invariant quantities that can be defined for asymptotically flat spacetimes. They consist of \emph{mass moments} $\mathscr{M}_i$, and \emph{current moments} $\mathscr{J}_i$ that uniquely characterize a space time. While mass moments describe how matter is distributed over the object, current moments describe the flow of matter. The original method proposed by Geroch and Hansen is quite laborious to implement, and we adopt a prescription by Fodor, Hohenselaers and Perjés \cite{Fodor:HP1989} using the Ernst potential outlined in\cite{Quevedo:1989rfm}.\footnote{an alternate way to compute this is described in detail in \cite{Vigeland:2010xe}.} The result is:
  \begin{align*}
	  \mathscr{M}_0 &= M+\frac{8}{3}a²Mq, & \mathscr{J}_0 &= 0,\\
	  \mathscr{M}_1 &= 0, & \mathscr{J}_1 &= aM,\\
	  \mathscr{M}_2 &= -a²M-\frac{2}{15}a²M³\left(16q-15p\right), & \mathscr{J}_2 &= 0.\\
  \end{align*}
  Higher moments are of the form $\mathscr{M}_{2k} = \left(-1\right)^k M a^{2k}$ and $\mathscr{J}_{2k+1} = \left(-1\right)^k M a^{2k+1}$ (which come from the Kerr metric) plus terms containing $p$ and $q$ that appear with $a^n$ where $n>2$. Since our exterior metric is a solution to Einstein's equations at order $a²$, the contribution from higher order terms cannot be trusted. The multipole structure of the exterior solution (up to the second moment)
  is therefore similar to the Kerr metric (i.e. $\mathcal{M}_n = M\left(i a\right)^n$ where $M_n \coloneqq \textrm{Re} \mathcal{M}_n$  and $J_n \coloneqq \textrm{Im} \mathcal{M}_n$) except for the mass which is shifted at order $a²$ and an additional quadrupole contribution. The mass shift is trivial and can be removed by adding to the above metric a perturbation of the form
  \begin{equation}
	  Δg_{μν}d x^μ d x^ν = -\frac{16}{3}a²Mq\left(\frac{r-M}{r²} d t²+\frac{r-M}{\left(r-2M\right)²} d r²+r d ϑ²+r\sin ²ϑ d φ²\right).
  \end{equation}
  This shifts $M_0$ back to $M$, leaving the rest of the moments unchanged. It can be checked that the metric in \eqref{eq:fullexternal}, with the above perturbation added to it, is a solution to Einstein's vacuum equations. This metric represents the space time outside a rotating object of mass $M$ and angular momentum $aM$. It has a quadrupole moment that is larger than a Kerr black hole by an amount proportional to $\left(16q-15p\right)$.
  
  \subsection{Inside the shell}
  Now that we have a solution on the exterior, let us try to find a solution describing the interior of the bubble.
  The AdS metric can be written in global coordinates as
  \begin{equation}
	  d s² = -\left(1+kr²\right) d t² + \left(1+kr²\right)^{-1} d r² + r² d ϑ² + r² \sin ²ϑ d φ²,
  \end{equation}
  To construct a generalization of AdS, let us make an ansatz of the form
  \begin{equation}
	  d s² = -\left(1+kr²\right)e^{μ_1} d t² + \left(1+kr²\right)^{-1}e^{μ_2} d r² + r² e^{μ_3} d ϑ² + r²\sin ²ϑ e^{μ_4} d φ²,
  \end{equation}
  where $μ_i$ are functions of $r$ and $ϑ$ in order to have a stationary axially symmetric solution. We choose the angular dependence of $μ_i$ to be proportional to the Legendre polynomial P$_2(\cos ϑ)\coloneqq \frac{1}{2}\left(3\cos ²ϑ-1\right)$ and a special relation between the $r$ dependent functions $f_i$ i.e.
  \begin{equation}\label{eq:insideHT}
	  \begin{split}
	  μ₁ &= f₁ \mathrm{P}_2\left(\cos ϑ\right),\\
	  μ₂ &= -f₁ \mathrm{P}_2\left(\cos ϑ\right),\\
	  μ₃ &= f₃ \mathrm{P}_2\left(\cos ϑ\right),\\
	  μ₄ &= f₃ \mathrm{P}_2\left(\cos ϑ\right).\\
	  \end{split}
  \end{equation}
  This separates out the angular dependence in Einstein's equations (with cosmological constant $Λ \equiv -3 k$) which can now be written as
  \begin{equation}
	  \begin{split}
	  \left(4+3kr²\right)f_1 -2f_3+r\left(1+kr²\right)f_1^{\prime} + r\left(3+4kr²\right)f_3^{\prime} + r²\left(1+kr²\right)f_3^{\prime\prime} &\overset{!}{=}0,\\
	  \left(3kr²-2\right)f_1 - 2f_3 + r\left(1+kr²\right)f_1^{\prime}+r\left(1+2kr²\right)f_3^{\prime} &\overset{!}{=}0,\\
	  2krf_1 + \left(1+kr²\right)\left(f_1^{\prime}+f_3^{\prime}\right) &\overset{!}{=}0,\\
	  6krf_1 + 2\left(1+3kr²\right)f_1^{\prime}+2f_3^{\prime}+4kr²f_3^{\prime}+r\left(1+kr²\right)\left(f_1^{\prime\prime}+f_3^{\prime\prime}\right)&\overset{!}{=}0,
	  \end{split}
  \end{equation}
  where primes denote derivatives with respect to $r$.
  We are looking for solutions to the above system of equations which vanish at the origin of the coordinate system.  
  Imposing this boundary condition we can pick the following solutions.\footnote{in a holographic language, going to the $3$d CFT on the boundary of AdS$_4$, one would expect a normalizable solution that goes like $1/r³$ and a non-normalizable solution which approaches a constant on the boundary. The solution here is a linear combination of these solutions with coefficients such that it vanishes at the origin.}
  \begin{equation}
	  \begin{split}
	  f₁ &= \frac{5 k r^2+3}{4 k^3 r^4+4 k^2 r^2} -\frac{3 \left(k r^2+1\right) \mathscr{P}}{4 k^{5/2} r^3},\\
	  f₃ &= \frac{4 k r^2-3}{4 k^2 r^2} -\frac{3\left(k r^2-1\right) \mathscr{P}}{4 k^{5/2} r^3},\\
	  \end{split}
  \end{equation}
  where $\mathscr{P}\coloneqq \arctan\left(\sqrt{k} r\right)$.
  
  Next, let us consider another axially symmetric generalization of the AdS metric of the form
  \begin{equation}
	  d s² = -\left(1+kr²\right)e^{σ_1} d t² + \left(1+kr²\right)^{-1}e^{σ_2} d r² + r² e^{σ_3} d ϑ² + r²\sin ²ϑ e^{σ_4} d φ²,
  \end{equation}
  where $σ_i$ are functions of $r$ and $ϑ$. Let us now choose the angular dependence of $σ_i$ to be such that $σ_4+σ_3\sim \mathrm{P}_2(\cos ϑ)$ and $σ_4-σ_3\sim \sin ²ϑ$. This is chosen so that the angular dependence separates out in Einstein's equations.\footnote{if we interchange the angular dependence of $σ₃$ and $σ₄$, the equations still separate but the only solution is $f₁=0, f₃=\textrm{constant}$, which corresponds to a constant shift in $ϑ$.} The radial parts for $σ_3$ and $σ_4$ are taken to be the same while the radial part of $σ_2$ is chosen to be proportional to $rg_3^{\prime}$.
  \begin{equation}\label{eq:insideour}
	  \begin{split}
	  σ₁ &= g₁ \left(3\cos ²ϑ -1\right),\\
	  σ₂ &= \frac{r}{6} g₃^{\prime}\left(3\cos ²ϑ -1\right),\\
	  σ₃ &= g₃ \cos ²ϑ,\\
	  σ₄ &= g₃ \left(2\cos ²ϑ -1\right).
	  \end{split}
  \end{equation}
  With this ansatz, the angular dependence separates out in Einstein's equations as designed and the equations take a simple form
  \begin{equation}
	  \begin{split}
	  6g₁+r\left(3+4kr²\right)g₃^{\prime}+r²\left(1+kr²\right)g₃^{\prime\prime} &\overset{!}{=} 0,\\
	  6g₁-\frac{r}{3}\left[6\left(1+4kr²\right)g₁^{\prime}+2\left(1+2kr²\right)g₃^{\prime}+6r\left(1+kr²\right)g₁^{\prime\prime}+r\left(2+kr²\right)g₃^{\prime\prime}\right] &\overset{!}{=}0,\\
	  6g₁+4g₃-2r\left(1+kr²\right)g₁^{\prime}-\frac{r}{3}\left(2+3kr²\right)g₃^{\prime} &\overset{!}{=}0,\\
	  4g₃-\frac{2}{3}r\left[2\left(1+2kr²\right)g₃^{\prime}+r\left(1+kr²\right)g₃^{\prime\prime}\right] &\overset{!}{=}0,\\
	  6g₁-6r\left(1+kr²\right)g₁^{\prime}-r\left(4+3kr²\right)g₃^{\prime} &\overset{!}{=}0,
	  \end{split}
  \end{equation}
  where primes denote derivatives with respect to $r$. As before, requiring that the solution to the above system of equations vanishes at the origin gives the following solution.
  \begin{equation}
	  \begin{split}
	  g₁ &= -\frac{2 k^2 r^4-2 k r^2-3}{6 k^3 r^4+6 k^2 r^2}-\frac{\mathscr{P}}{2 k^{5/2} r^3},\\
	  g₃ &= \frac{k r^2-3}{3 k^2 r^2}+\frac{\mathscr{P}}{k^{5/2} r^3},
	  \end{split}
  \end{equation}
  where $\mathscr{P}$ is defined as before.
  The two solutions obtained above can be combined to describe the spacetime inside the shell
  \begin{equation}\label{eq:interiorcombined}
	  \begin{split}
	  d s² =& -\left(1+k r²\right) e^{a²\left(c₁μ₁+c₂σ₁\right)} d t² + \left(1+k r²\right)^{-1} e^{a²\left(c₁μ₂+c₂σ₂\right)} d r²\\
	  & + r² e^{a²\left(c₁μ₃+c₂σ₃\right)} d ϑ² + r² \sin ²ϑ e^{a²\left(c₁μ₄+c₂σ₄\right)} d φ²,
	  \end{split}
  \end{equation}
  where $μ_i$ and $σ_i$ are given by \eqref{eq:insideHT} and \eqref{eq:insideour} respectively with $c_i$ being arbitrary constants.
  
  Having set up the geometries inside and outside the bubble, we now need to match them across the shell by imposing continuity of the induced metric. We begin by working at first order in $a$, which serves to outline the main idea behind our approach. Subsequently, we present the second order computation in section \ref{sec:secondorder}.
  
  \section{First order in spin}\label{sec:firstorder}
  To lowest order in $a$, the exterior metric from the previous section reduces to
  \begin{equation}\label{eq:firstorderKerr}
	  \begin{split}
	  d s_+² = -&\left(1-\frac{2M}{r}\right) d t² + \left(1-\frac{2M}{r}\right)^{-1} d r² + r² d ϑ² + r² \sin ²ϑ d φ²\\
	  &+\frac{4aM}{r} \sin ²ϑ d t d φ,
	  \end{split}
  \end{equation}
  while in this limit, the metric in the interior described by \eqref{eq:interiorcombined} is pure AdS i.e.
  \begin{equation}\label{eq:firstorderads}
	  d s_-² = -\left(1+k \tilde{t}²\right)d \tilde{t}² + \left(1+k \tilde{r}²\right)^{-1} d \tilde{r}² + \tilde{r}² d \tilde{ϑ}² + \tilde{r}² \sin ²\tilde{ϑ} d \tilde{φ}²,
  \end{equation}
  where $\left(t,r,ϑ,φ\right)$ are coordinates outside the shell, and $\left(\tilde{t},\tilde{r},\tilde{ϑ},\tilde{φ}\right)$ are coordinates in the interior of the shell.
  
  In the absence of rotation, i.e. $a=0$, this is just the Schwarzschild case discussed in \cite{Danielsson:2017riq} with the shell positioned at $R=9M/4$. For non-zero spin ($a\neq 0$), radius of the shell gets corrected at order $a²$, i.e. $r=R+\mathcal{O}\left(a²\right)$, where $R\coloneqq 9M/4$.\footnote{the radius cannot have corrections at odd powers of $a$ since it cannot depend on the sign of $a$, i.e., the direction in which the shell spins.}
  
  The metric induced on the shell $Σ$ from outside is given by
  \begin{equation}
	  d s²_{Σ_+} = -\left(1-\frac{2M}{R}\right) d t² + R² d ϑ² + R² \sin ²ϑ d φ² + \frac{4aM}{R} \sin ²ϑ d t d φ,
  \end{equation}
  which can be matched to the metric induced from inside by rotating the angular coordinate to $\tilde{φ} = φ + a Ω t$, and rescaling time to $\tilde{t} = A t$, where
  \begin{equation}\label{eq:omegaAsol}
	  Ω = \frac{2aM}{R},\qquad \qquad A = \sqrt{\frac{1-2M/R}{1+kR²}}.
  \end{equation}
  This is the first junction condition, which ensures that the induced metric is continuous across the shell.
  The stress-energy tensor on the shell is given by the jump in extrinsic curvature across it i.e.
  \begin{equation}\label{eq:stressenergytensor}
	  S_{ab} = -\frac{1}{8π}\left(\left[K_{ab}\right]-\left[K\right]h_{ab}\right),
  \end{equation}
  where $\left[ \cdot \right]$ is the jump of the corresponding quantity across the shell, $K_{ab}$ is the extrinsic curvature, $K$ is its trace, and $h_{ab}$ is the induced metric.
  This gives
  \begin{equation}\label{eq:firstorderstressenergy}
	  \begin{split}
	  S^t _t &= \frac{1}{4π R}\left(\sqrt{1-\frac{2M}{R}}-\sqrt{1+kR²}\right),\\
	  S^ϑ _ϑ = S^φ _φ &= \frac{1}{8πR}\left(\frac{1-M/R}{\sqrt{1-2M/R}}-\frac{1+2kR²}{\sqrt{1+kR²}}\right),\\
	  S^t _φ &= -\frac{1}{8 \pi R²}\frac{3aM \sin ² ϑ}{\sqrt{1-2M/R}},\\
	  S^φ _t &= \frac{M a}{8\pi R^4} \left(\frac{2}{\sqrt{1+kR²}}+\frac{1}{\sqrt{1-2M/R}}\right).
	  \end{split}
  \end{equation} 
  Using $R \coloneqq 9M/4$, the components of the stress-energy tensor up to leading orders in $k$ become
  \begin{equation}\label{eq:totalstressenergyfirstorder}
	  \begin{split}
	  S^t _t &= -\frac{√k}{4π}+\frac{1}{27Mπ}-\frac{2}{81√kM²π},\\
	  S^ϑ _ϑ &= -\frac{√k}{4π} + \frac{5}{54Mπ},\\
	  S^φ _φ &= -\frac{√k}{4π} + \frac{5}{54Mπ},\\
	  S^t _φ &= -\frac{2a \sin ²ϑ}{9Mπ},\\
	  S^φ _t &= \frac{32a}{2187M³π}+\frac{256a}{59049√k M⁴π},
	  \end{split}
  \end{equation}
  Let us now try to cast this in the form of a perfect fluid.
  The stress-energy tensor of a perfect fluid is given by
  \begin{equation}
	  S^i _j = \left(ρ + p\right) u^i u_j + p δ^i _j,
  \end{equation}
  where $ρ$ and $p$ are the density and pressure of the fluid, while $u^i$ is its velocity vector. Since the shell is made of high tension branes (with equation of state $p=-ρ$), a gas of open strings (with equation of state $p=ρ/2$) sitting on top of it and stiff matter (with equation of state $p=ρ$) made of D0 branes, the stress-energy tensor can be written as a sum of these three components. The total stress energy tensor in \eqref{eq:totalstressenergyfirstorder} should therefore be split into $S_{\textrm{total}}=S_{\textrm{brane}}+S_{\textrm{gas}}+S_{\textrm{stiff}}$, where
  \begin{equation}\label{eq:firstorderSsplit}
	  \begin{split}
	  \left(S_{\textrm{brane}}\right)^i _j &= \left(-\frac{√k}{4π}+\frac{2}{27Mπ}-\frac{1}{81√kM²π}\right)δ^i _j,\\
	  \left(S_{\textrm{gas}}\right)^i _j &= \frac{1}{18Mπ}u^i u_j+\frac{1}{54Mπ}δ^i _j,\\
	  \left(S_{\textrm{stiff}}\right)^i _j &= \frac{2}{81√kM²π}u^i u_j+\frac{1}{81√kM²π}δ^i _j.
	  \end{split}
  \end{equation}
  The velocity vector $u^i$ should be of the form $u^i \equiv \left(γ,0,a β\right)$ where $β$ corresponds to the rotation. Comparing with the stress-energy tensor and normalizing ($u^i u_i \overset{!}{=}-1$), we get 
  \begin{equation}\label{eq:ufirstorder}
	  u^i \equiv \left(3, 0, -a \frac{64}{243M²}\frac{27 \sqrt{k} M+8}{9 \sqrt{k} M+4}\right).
  \end{equation}
  
  \section{Second order in spin}\label{sec:secondorder}
  Now working at order $a²$, the external metric has the following components
  \begin{equation}
	  \begin{split}
	  \tilde{g}_{tt} &= g_{tt}\left(1+2 q a² λ + 2 p a² χ\right),\\
	  \tilde{g}_{rr} &= g_{rr}\left(1+2 q a² ν - 2 p a² χ\right),\\
	  \tilde{g}_{ϑϑ} &= g_{ϑϑ}\left(1+2 q a² ν + 2 p a² ψ\right),\\
	  \tilde{g}_{φφ} &= g_{φφ}\left(1-2 q a² λ + 2 p a² ψ\right),\\
	  \tilde{g}_{tφ} &= g_{tφ},
	  \end{split}
  \end{equation}
  where $g_{μν}$ are defined in \eqref{eq:kerr_unperturbed}. 
  The metric on the inside is given by \eqref{eq:interiorcombined} i.e.
  \begin{equation}
	  \begin{split}
	  g_{tt} &= \left(1+k r²\right)\left(1+ a² \left(c₁ μ₁+c₂ σ₁\right)\right),\\
	  g_{rr} &= \left(1+k r²\right)^{-1}\left(1+ a² \left(c₁μ₂+c₂σ₂\right)\right),\\
	  g_{ϑϑ} &= r²\left(1+ a²\left(c₁μ₃+c₂σ₃\right)\right),\\
	  g_{φφ} &= r² \sin ²ϑ\left(1+ a² \left(c₁μ₄+c₂σ₄\right)\right),
	  \end{split}
  \end{equation}
  The functions $λ,ν,χ,ψ,μ_i$ and $σ_i$ are defined in \eqref{eq:oursols}, \eqref{eq:HTsols}, \eqref{eq:insideHT} and \eqref{eq:insideour}. Generically, the radius of the shell is no longer a constant but receives $\mathcal{O}\left(a²\right)$ corrections. We parametrize the radius in terms of external coordinates $\left(t,r,ϑ,φ\right)$ and internal coordinates $\left(\tilde{t},\tilde{r},\tilde{ϑ},\tilde{φ}\right)$ as
  \begin{equation}
	  \begin{split}
	  r &= R + R a²\left(m_1 + m_2 \cos 2ϑ\right),\\
	  \tilde{r} &= R+R a²\left(n_1 + n_2 \cos 2ϑ \right),
	  \end{split}
  \end{equation}
  for arbitrary constants $m_i$ and $n_i$. Similar to the first order computation, we impose continuity of the induced metric across the shell by rescaling the time coordinate $\tilde{t}$ in terms of $t$ and introducing a rotation $\tilde{φ}=φ + Ω t$. 
  This determines $m_1,m_2,n_2,c_1$ and $c_2$, leaving three undetermined parameters $n_1,p$ and $q$.
  The first order expression for $A$ in \eqref{eq:omegaAsol} receives a correction at order $a²$. Solutions to this first junction condition are listed in appendix \ref{app:junctionconditions}.
  
\begin{figure}[t]
  	\centering
  	\begin{subfigure}[t]{0.45\textwidth}
  		\includegraphics[width=0.9\textwidth]{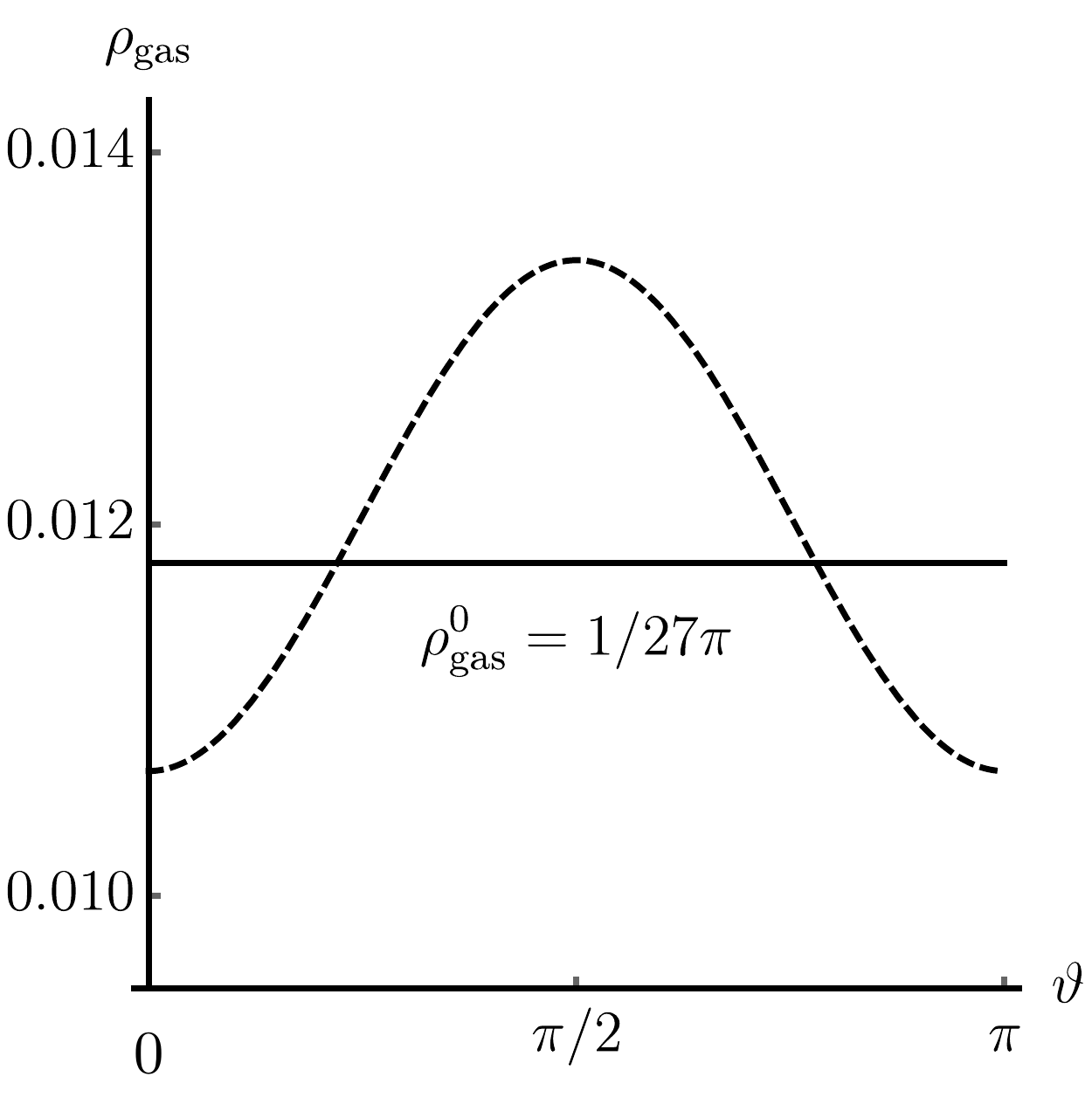}
  		\caption{\emph{Density of the string gas}}
  		\label{fig:density}
  	\end{subfigure}
  	~
  	\begin{subfigure}[t]{0.45\textwidth}
  		\includegraphics[width=0.9\textwidth]{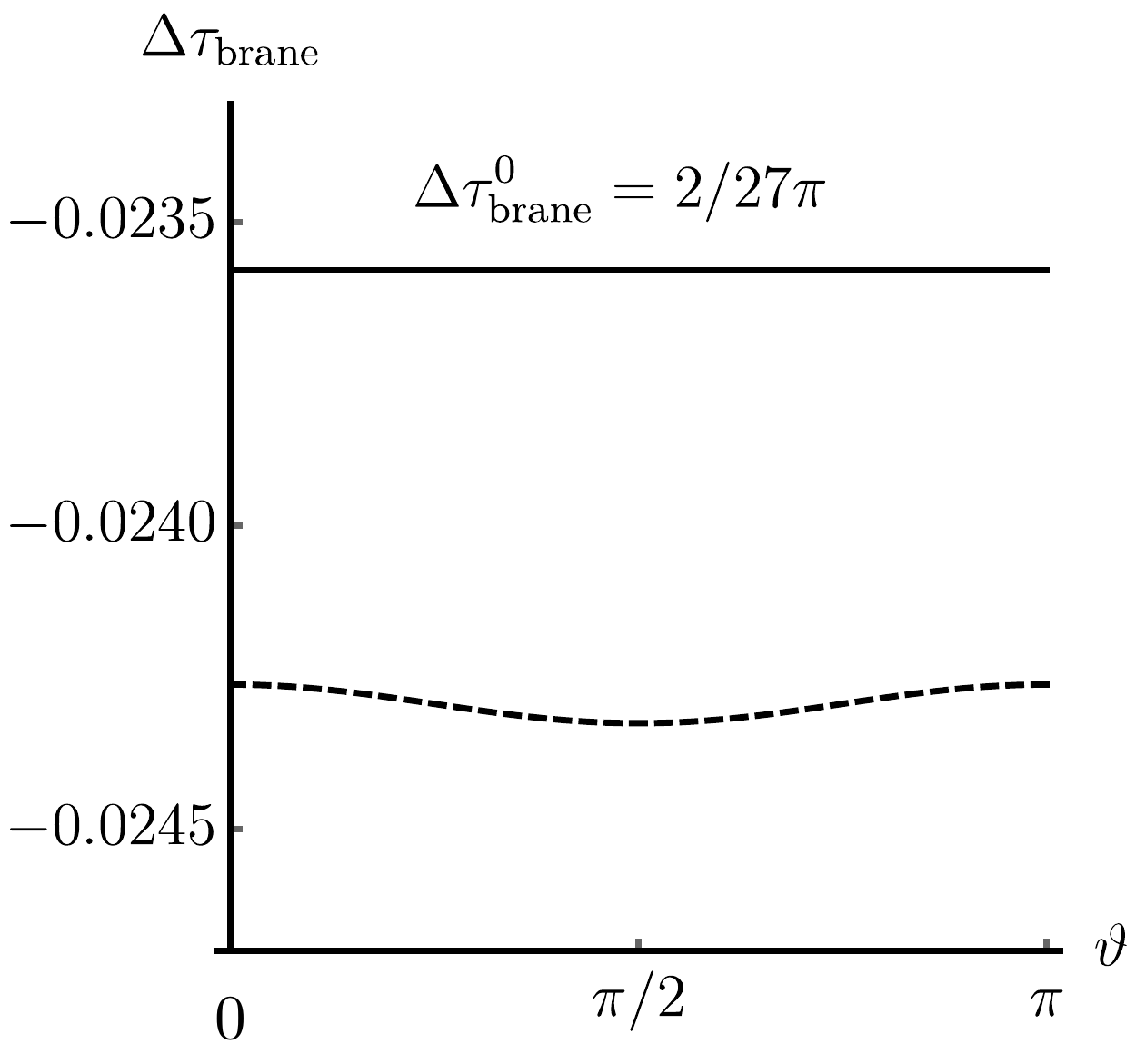}
  		\caption{\emph{Tension of the shell}}
  		\label{fig:tension}
  	\end{subfigure}
  	\caption{\emph{Density of the gas and tension of the shell (apart from the $√k/4π$ contribution) as a function of the angular variable $ϑ$ \emph{(}for $M=1, a=0.3$\emph{)} are plotted as dashed lines. Density is given by $ρ_{\textrm{\emph{gas}}}=0.0118-a^2 (0.0212+0.106 n_1+0.0153\cos 2ϑ)$ while the tension $Δτ_{\textrm{\emph{brane}}}$ is given by $Δτ_{\textrm{\emph{brane}}} \coloneqq τ-\sqrt{k}/4π = -0.0236 + a² \left(0.00812 + 0.0707 n_1 +0.000354 \cos 2ϑ\right)$.  The straight lines show the corresponding values for a non-rotating shell. The density of the gas (and correspondingly the pressure) as well as the pressure due to the brane $(p=-ρ)$ increases towards the equator causing a net increase in the pressure which provides the necessary centripetal force that holds the bubble together as it starts to spin.}}\label{fig:densityandtension}
  \end{figure}

  The stress-energy tensor on the shell can now be computed from the jump in the extrinsic curvature across the shell i.e. \eqref{eq:stressenergytensor}. We find
  \begin{equation}\label{eq:stressenergysecondorder}
	  \begin{split}
	  S^t _t &= -\frac{√k}{4π}
	  + \frac{1}{27Mπ} - \frac{2}{81√kM²π} + a²\left(\mathcal{X}_1 + \mathcal{Y}_1\cos 2ϑ+\frac{1}{√k}\left(\mathcal{H}_1+\mathcal{K}_1\cos 2ϑ\right)\right),\\
	  S^ϑ _ϑ &= -\frac{√k}{4π}
	  + \frac{5}{54Mπ} + a²\left(\mathcal{X}_2 + \mathcal{Y}_2\cos 2ϑ+\frac{1}{√k}\left(\mathcal{H}_2+\mathcal{K}_2\cos 2ϑ\right)\right),\\
	  S^φ _φ &= -\frac{√k}{4π}
	  + \frac{5}{54Mπ} + a²\left(\mathcal{X}_3 + \mathcal{Y}_3\cos 2ϑ+\frac{1}{√k}\left(\mathcal{H}_3+\mathcal{K}_3\cos 2ϑ\right)\right),\\
	  S^t _φ &= -\frac{2a \sin ²ϑ}{9Mπ},\\
	  S^φ _t &= \frac{32a}{2187M³π}+\frac{256a}{59049√k M⁴π},
	  \end{split}
  \end{equation}
  where the quantities $\mathcal{X}_i$, $\mathcal{Y}_i, \mathcal{H}_i$ and $\mathcal{K}_i$ are functions of $p,q$ and $n_1$ (see appendix \ref{app:stressenergytensor}).
  For the shell to be made of branes with a gas of massless open strings and stiff matter sitting on top, the stress-energy tensor should be writable in the form $S_{\textrm{total}}=S_{\textrm{brane}}+S_{\textrm{gas}}+S_{\textrm{stiff}}$ as before. This determines the constants $p$ and $q$ of the metric uniquely as
  \begin{equation}\label{eq:pqnumbers}
	  p = -\frac{0.144}{M²}+\mathcal{O}\left(1/√k\right),\qquad \qquad 
	  q = \frac{0.0162}{M²}+\mathcal{O}\left(1/√k\right),
  \end{equation}
  and the stress-energy tensor splits into\footnote{arguing like in \cite{Danielsson:2017riq} that the gas does not have $k^{-1/2}$ pieces, which is split between the stiff gas and the branes.}
  \begin{equation}\label{eq:stressenergysplit}
	  \begin{split}
	  \left(S_{\textrm{brane}}\right)^i _j &= \left(-\frac{√k}{4π} + \frac{2}{27Mπ} -\frac{1}{81√kM²π} + a²\mathcal{Z}_1\right)δ^i _j,\\
	  \left(S_{\textrm{gas}}\right)^i _j   &= \left(\frac{1}{18Mπ} + a² \mathcal{Z}_2\right)u^i u_j +
	  \left(\frac{1}{54Mπ} + \frac{a²}{3} \mathcal{Z}_2\right) δ^i _j,\\
	  \left(S_{\textrm{stiff}}\right)^i _j &= \left(\frac{2}{81√kM²π} + a² \mathcal{Z}_3
	  \right)u^i u_j+\left(\frac{1}{81√kM²π}+ \frac{a²}{2} \mathcal{Z}_3\right)δ^i _j,
	  \end{split}
  \end{equation}
  where $\mathcal{Z}_i$ are functions involving constants and $n_1$ (see appendix \ref{app:stressenergytensor}). The velocity vector $u^i$ now has corrections of order $a²$ over the first order result in \eqref{eq:ufirstorder} (see appendix \ref{app:stressenergytensor}). We see that the stress-energy tensor from the first order computation in \eqref{eq:firstorderSsplit} gets corrected at order $a²$ just as expected.
  We further notice that the angular dependence of the radius (governed by $n_2$) goes as $\mathcal{O}\left(k^{-1}\right)$ to leading order in $k$. This means that the shell is spherical for large $k$.
  The constants $m_1$ and $n_1$ are not determined by the junction conditions, and we have to make a further physical argument to fix the radius of the shell. A simple possibility is to assume that the total amount of fluid on the shell (which includes the gas and stiff matter) is conserved when the shell starts to rotate i.e.
  \begin{equation}
	  4π r²_{\textrm{Schw}} ρ_{\textrm{Schw}} = \int\limits_{0}^{π}\D ϑ\int\limits_{0}^{2π}\D φ ρ_{\textrm{rot}}r²_{\textrm{rot}}\sin ϑ,
  \end{equation}
  where $r_{\textrm{Schw}}$ and $ρ_{\textrm{Schw}}$ are the radius and density of the fluid on the stationary black shell while $r_{\textrm{rot}}$ and $ρ_{\textrm{rot}}$ are the corresponding quantities for the rotating shell. This
  determines 
  $n_1=-0.232/M²+\mathcal{O}\left(1/√k\right)$, which fixes the radius at
  \begin{equation}
	  r = \frac{9M}{4}-\frac{0.81a²}{M}+\mathcal{O}\left(1/√k\right),
  \end{equation}
  up to leading order in $k$ and we find that the shell shrinks a little. The density of the gas and the tension of the shell are shown in figure \ref{fig:densityandtension}.
  
  It is interesting to note that although the radius of the shell is not fully determined by the junction conditions without further physical input beyond the equations of state, the quadrupole moment is fixed. It is independent of $n_1$ and is \emph{uniquely} given by
  \begin{equation}\label{eq:quadrupole}
	  \begin{split}
	  Q &= -a²M-\frac{2}{15}a²M³\left(16q-15p\right)\\
	  &= -1.32 a²M,
	  \end{split}
  \end{equation}
  which is an increase of about $32\%$ compared to that of a Kerr black hole with the same spin.
  
  We would like to point out two things here. Firstly,
  it is possible that energy is exchanged between the branes and the massless gas once the shell starts to spin. This would give a slightly different value for $n_1$ and consequently $r$. Details of the string theory construction of the shell would give us the exact dynamics of these components and thus a way to fix the quantity $n_1$. However, since the observable of interest i.e. the quadrupole moment does not depend on $r$, we do not explore this further.
  
  Secondly, it should also be noted that we have picked specific solutions to Einstein's equations both in the interior and in the exterior of the shell that we then used to solve the junction conditions. These are not the only possible solutions, but none of the other solutions we investigated solved the junction conditions, given the physical properties of the shell that we assume. This indicates that our choice is the correct physical one. One would also expect on physical grounds that the metric outside the spinning shell is unique up to coordinate transformations.
  
  One can compare this with the case of a compact object
  such as a neutron star, where the exterior metric is not expected to be described by the Kerr metric. The deviations from the Kerr metric, including a different quadrupole moment, are determined by the physical properties of the star.
  
  \section{Astrophysical implications}\label{sec:astrophysical}
  
  Let us now discuss some observational implications of our proposal.
  Our model has no free parameters, and predicts specific observational signatures such as a significant increase in the quadrupole moment of about $32\%$ compared to that of a Kerr black hole with the same spin. 
  Our results are valid for small spins, but it is reasonable to speculate that the quadrupole moment would differ from that of the Kerr solution also at moderately large spins.  
  This, however, might change when the spin approaches its maximal value. If the shell then approaches the would be horizon (as it does when the charge of a Reissner-Nordström black hole is increased towards extremality \cite{Danielsson:2017riq}), the no-hair theorem could be restored, which would imply that the shift in the quadrupole moment vanishes as $a\rightarrow 1$. We hope to generalize our results to arbitrary values of $a$ in a future work.
  Although the spins of some  black holes have been measured (see for example \cite{Risaliti:2013cbe,Reis:2014ega}), it has, so far, not been possible to accurately measure the quadrupole moment. Luckily, this might change soon and we will discuss a few possibilities below.
  
  An obvious possibility would be through high precision measurements of the gravitational radiation emitted by colliding black holes. LIGO studies colliding stellar mass black holes, and will be able to measure the quadrupole moment with an accuracy of the order of the Kerr moment for higher spins. This suggests that it is unlikely for LIGO to reach the sensitivity required to test our model.\footnote{e.g. it was shown in \cite{Konoplya:2016pmh} that the large indeterminacy in mass and angular momentum in LIGO and VIRGO measurements allows for rather large deviations from the Kerr geometry to be consistent with the gravitational wave ring down profile.} LISA, on the other hand, focuses on super massive black holes and will reach much higher sensitivities. An analysis of the sensitivity of LIGO and LISA was given in \cite{Krishnendu:2017shb} and it seems likely that measurements for LISA will be able to confirm or rule out our model.
  
  There are also other ways of constraining the quadrupole moments through astrophysical observations -- one of them being the study of accretion discs around black holes. Infalling matter can migrate inwards, converting gravitational energy into kinetic energy and radiation, until it reaches the innermost stable orbit \cite{Blandford:1982di}. If it keeps falling beyond this orbit, it is rapidly captured by the black hole and no energy is released. The outcome is the same irrespective of whether one deals with a black hole or a black shell. The fraction of infalling matter which is converted to radiation ($\coloneqq η$) can be used as a measure of this efficiency. While the luminosity of accretion discs is easy to measure, the accretion rate is much harder to obtain. For instance, the black hole at the center of the galaxy has an efficiency no larger than $η \sim 5 \cdot 10^{-6}$ \cite{Narayan:1997ku}. The theoretical limit for the efficiency can be calculated from the effective potential and increases from $η\sim 0.057$ for Schwarzschild to $η\sim 0.42$ for a maximally spinning Kerr black hole.
  
  The efficiency for our model is given by (up to order $a²$)
  \begin{equation}\label{eq:efficiency}
	  η=\underbrace{1-\frac{2 \sqrt{2}}{3}}_{\textrm{Schw.}}+\underbrace{\frac{a}{18 \sqrt{3}}+\frac{5 a^2}{162 \sqrt{2}}}_{\textrm{Kerr}}-\underbrace{\frac{a^2 \left(185-456 \log \left(\frac{3}{2}\right)\right) (16 q-15 p)}{72 \sqrt{2}}}_{\textrm{quadrupole}},
  \end{equation}
  where the corresponding pieces coming from Schwarzschild, Kerr and the quadrupole are identified above. 
  An increase in the combination $\left(16q-15p\right)$ results in an increase of the quadrupole moment. 
  Since the coefficient in front of this term in  \eqref{eq:efficiency} is negative, an \emph{increase} of the quadrupole moment results in a \emph{decrease} in efficiency. However, since this coefficient is suppressed by an order of magnitude over the Kerr and Schwarzschild pieces, the decrease in efficiency is very small for slowly spinning black shells ($\sim 0.05\%$ for $a=0.1$).
  Current estimates suggest $η>0.15$ \cite{Elvis:2001bn} as a lower limit and that a reasonable estimate for a mean value is $η\sim 0.30-0.35$ \cite{Wang:2006bz}. As noted in \cite{Bambi:2011uf}, the observation of a single object with a high value of $η (>\sim 0.42)$ would be enough to rule out the Kerr metric. The situation is the same for black shells.
  
  One should note that there are other mechanisms that power the Active Galactic Nucleus (AGN). Through the Blandford-Znajek (BZ) mechanism \cite{Blandford:1977ds}, the energy to power jets emanating from the AGN can be extracted from the rotational energy of the black hole. In this way the efficiency is no longer limited by the 
  inflow of matter into the black hole, and can even exceed $1$. The BZ mechanism is a version of the Penrose process \cite{Penrose:1971uk} and makes use of the ergosphere surrounding a spinning black hole. Particles inside the ergosphere can have energies that are negative as measured from far away. In the Penrose process, a particle that enters the ergosphere splits into two, one of which has a negative energy and is captured by the black hole, while the other escapes with higher energy than the incoming one. In this way energy can be extracted from a spinning black hole without violating the laws of black hole thermodynamics 
  since the reduction in energy is accompanied by a reduction in spin such that the area of the horizon increases. The BZ mechanism works in a similar way but makes use of electromagnetic fields instead of particles.
  
  The reason why an ergosphere is necessary in order to extract rotational energy by the above processes, is because of the nature of the black hole horizon. With an ordinary rotating object, one can extract energy by scattering particles against the rotating surface. If the particle hits the surface 
  with a speed lower than the speed of rotation, it can simply scatter off with an increased speed and energy. However, if the rotating object is a black hole, such a scattering is not possible since there is no surface to scatter from and the particle is sucked into the horizon resulting in an increase rather than a decrease in the energy of the black hole.
  
  It works in the same way for black shells even though there is no horizon. As we have argued, the surface of the shell appears completely black due to the large number of degrees of freedom that effectively absorb anything that is incident on the shell (since it is entropically favorable). As a consequence, scattering off the shell is again of limited use if one wants to extract rotational energy from the shell. Hence, the use of an ergosphere is essential also in the case of a black shell. Since the shell starts out far outside of where the horizon  (or even the ergosphere) would have been positioned, there is no ergosphere for a slowly spinning black shell. As the spin increases, we expect the shell to shrink in size. If there is an analogue to the case of charged black holes discussed in \cite{Danielsson:2017riq}, we might expect the shell to approach the would be horizon as the spin becomes maximal. If this is true, there should exist a critical value of the spin $a$ such that for higher values, an ergosphere (or a part of it) starts to appear outside the shell. A fully developed ergosphere is then expected to appear only in the limit of maximal spin. This should affect the efficiency of the BZ-mechanism as a function of the spin.
  
  In this context one needs to address the possibility of an {\it ergosphere instability}. It was argued in \cite{friedman1978} that any object with an ergosphere, but without a horizon, is unstable. However, as discussed in \cite{Maggio:2017ivp}, the essential feature of the horizon that removes the instability is that it effectively absorbs all incoming negative energy states. As the authors show, a compact object with an absorption coefficient as small as $1\%$ can escape the instability. For our black shells, which are expected to be as efficient absorbers as real black holes, the ergosphere instability is therefore not an issue.
  
  A promising approach to measure the quadrupole moment with sufficient accuracy, is through the study of stars in close orbits around a black hole e.g. the black hole at the center of the Milky Way, coinciding with the radio source Sgr A*. For a review, see \cite{Johannsen:2015mdd}. The orbit of a star will precess in the plane of the orbit, with frame dragging due to the spin as the dominating effect. In addition, the angular momentum of the orbit will precess with a rate that scales with the quadrupole moment. With the use of GRAVITY, an interferometer on the Very Large Telescope (VLT), one hopes to be able to find, and study, stars with semi major axis of the order $1000 \textrm{GM}/\textrm{c}²$, and high eccentricity \cite{Psaltis:2015uza}. With two stars, and if they are bright enough, one can reach a sensitivity that could distinguish a black shell from a black hole. Even better, if one could find a pulsar with an orbital period of less than half a year, one should easily be able to exclude the existence of a black shell, and confirm the no-hair theorem with high accuracy.
  
  Another possibility, which  could be close in time, is to study the shadow of the Sgr A* black hole using the Event Horizon Telescope. The black hole shadow has an asymmetry of order $a^3$, and is therefore symmetric for all but the fastest spinning Kerr black holes. With a non-Kerr quadrupole moment, an extra term appears with the asymmetry of order $a^2$, providing a way to measure deviations from the Kerr metric. Again, the expected accuracy is such that it might just be possible to distinguish a black shell from a black hole.
  
  There could also be other kinds of signals that could be used to distinguish between black shells and black holes. For instance, the presence of a hard surface outside of the horizon could lead to modifications of the gravitational waves emitted by colliding black holes, as discussed in \cite{Abedi:2016hgu,Cardoso:2016oxy,Cardoso:2016rao,Barcelo:2017lnx}.
  
  To summarize, there are plenty of opportunities within the next few years to constrain models that replaces true black holes with horizonless structures such as the black shells.
  
  \section{Conclusions}\label{sec:conclusions}

  In this paper we have built on the proposal of \cite{Danielsson:2017riq} where it was suggested that the end point of gravitational collapse is not a black hole but rather a thin shell (enclosing a bubble of AdS space with a large negative cosmological constant) made out of branes from string theory with other stringy matter on top.
  We have generalized the construction to slowly rotating black shells that mimic black holes with low angular momentum. We have constructed explicit metrics describing space time in the interior and exterior of such shells, and have solved the junction conditions to find the shape of such objects. We found that the shells are spherical up to leading order in the cosmological constant $k$, even though the junction conditions allow for an indeterminacy of the radius at order $a²$. Further physical input is needed beyond the equations of state, and we suggest that the shell will actually shrink when it starts to rotate. Curiously enough, the quadrupole moment does not depend on this indeterminacy and is unique. It turns out to be about $32\%$ larger than the corresponding value for a Kerr black hole with the same spin. We propose the modified value of the quadrupole moment as a test of our model, and expect that it should be possible to measure it with sufficient accuracy in the not too distant future. Promising methods include imaging of the black hole at the center of the Milky Way using the Event Horizon Telescope, precision measurements of stars in close orbits around the central black hole, and future observations of colliding super massive black holes using LISA. In order to make contact with the quadrupole moment of fast spinning black holes that might be observed astrophysically, the present work needs to be extended for arbitrary values of $a$.
  
  An important feature of spinning black holes, as well as a spinning black shells, is the ergosphere, which allows for the extraction of rotational energy through the Blandford-Znajek mechanism. For a slowly rotating black shell there is no ergosphere, but we have argued that it is expected to appear at some critical value of the spin. To understand these aspects better, our analysis needs to be extended to general values of the spin parameter $a$. We hope to return to this important, but challenging, calculation in a future work. Additionally, similar to the non-rotating black shells in \cite{Danielsson:2017riq}, we hope to check the perturbative stability, thermodynamic properties and formation mechanism of these rotating black shells in a future work.
  
  \section*{Acknowledgements}
  We would like to thank Marek Abramowicz, Souvik Banerjee and Giuseppe Dibitetto for stimulating discussions. We would also like to thank Giuseppe Dibitetto for going through this manuscript and providing detailed feedback. The work of the authors was supported by the Swedish Research Council (VR).
	\appendix
	\section{Solutions to the first junction condition}\label{app:junctionconditions}
	Solutions to the first junction condition are obtained by matching the induced metric across the shell. Since the full solution (to all orders in $k$) is unwieldy, here we present the results only up to order $k^{-1/2}$.
	\begin{align*}
		A =& \frac{4}{3\sqrt{81kM²+16}}+δA,\\
		δA =& \frac{1}{√k M³}\left(\frac{4 M^2 n₁}{9}-\frac{5 M^2 p (100+243 \log 3)}{15552}-\frac{1}{324} M^2 q (265 \log 3-148)+\frac{2384}{59049}\right),\\
		Ω =& \frac{128}{729M²},\\
		m₁ =& -\frac{208}{2187 M^2}+n₁+\frac{1}{432} q (99 \log 3-604),\\
		m₂ =& \frac{172}{2187 M^2}+\frac{5 p (100+243 \log 3)}{2304}+\frac{1}{24} q (101 \log 3-68),\\
		n₂ =& \frac{172}{2187 M^2}+\frac{5 p (100+243 \log 3)}{2304}+\frac{1}{48} q (265 \log 3-148),\\
		c₁ =& \frac{5 p (1323 \log 3-1276) \left(36 k M^2+6 \pi  \sqrt{k} M+\pi ^2\right)}{10368 M^2}+\frac{2 q (5 \log 3-4) \left(36 k M^2+6 \pi  \sqrt{k} M+\pi ^2\right)}{27 M^2}\\
		&+\frac{208 \left(36 k M^2+6 \pi  \sqrt{k} M+\pi
			^2\right)}{19683 M^4},\\
		c₂ =& -\frac{4 q (5 \log 3-4) \left(27 k M^2+16\right)}{9 M^2}-\frac{272 \left(27 k M^2+16\right)}{6561 M^4}.
	\end{align*}
	Evaluating them numerically gives
	\begin{align*}
		A =& \frac{0.148}{√kM}+δA,\\
		δA =& \frac{1}{√kM}\left(0.444n₁-0.118p-0.442q+\frac{0.0404}{M²}\right),\\
		Ω =& \frac{0.176}{M²},\\
		m₁ =& -\frac{0.0951}{M^2}+n₁-1.15 q,\\
		m₂ =& \frac{0.0786}{M^2}+0.796 p+1.79 q,\\
		n₂ =& \frac{0.0786}{M^2}+0.796 p+2.98 q,\\
		c₁ =& \left(\frac{0.199 \sqrt{k}}{M^3}+\frac{0.380 k}{M^2}+\frac{0.104}{M^4}\right)+p \left(\frac{1.61 \sqrt{k}}{M}+3.08 k+\frac{0.845}{M^2}\right)\\
		& +q \left(\frac{2.08 \sqrt{k}}{M}+3.98 k+\frac{1.09}{M^2}\right),\\
		c₂ =& -\frac{1.12 k}{M^2}-\frac{0.663}{M^4}+q \left(-17.9 k-\frac{10.6}{M^2}\right).\\
	\end{align*}
	
	\section{Stress energy tensor on the shell}\label{app:stressenergytensor}
	Stress energy tensor on the shell can be computed from the jump in the induced metric across the shell using \eqref{eq:stressenergytensor}. This gives \eqref{eq:stressenergysecondorder} up to $\mathcal{O}\left(k^{-1/2}\right)$ where
	the quantities $\mathcal{X}_i$ and $\mathcal{Y}_i$ are given by
	\begin{align*}
		\mathcal{X}₁ =& \frac{2187 M^2 n₁+380}{19683 \pi  M^3}+\frac{5 p (1053 \log 3-932)}{10368 \pi  M}+\frac{32 q (3 \log 3-2)}{81 \pi  M},\\
		\mathcal{Y}₁ =& \frac{2080}{19683 \pi  M^3}+\frac{5 p (6561 \log 3-5492)}{20736 \pi  M}+\frac{q (1801 \log 3-1172)}{432 \pi  M},\\
		\mathcal{X}₂ =& -\frac{15309 M^2 n₁+932}{39366 \pi  M^3}+\frac{35 p (100+243 \log 3)}{41472 \pi  M}+\frac{q (1065 \log 3-788)}{648 \pi  M},\\
		\mathcal{Y}₂ =& -\frac{176}{19683 \pi  M^3}+\frac{5 p (100+243 \log 3)}{20736 \pi  M}+\frac{q (156-307 \log 3)}{864 \pi  M},\\
		\mathcal{X}₃ =& -\frac{15309 M^2 n₁+1028}{39366 \pi  M^3}+\frac{5 p (100+243 \log 3)}{41472 \pi  M}+\frac{2 q (4+3 \log 3)}{81 \pi  M},\\
		\mathcal{Y}₃ =& -\frac{128}{19683 \pi  M^3}+\frac{5 p (100+243 \log 3)}{10368 \pi  M}+\frac{q (1049 \log 3-980)}{864 \pi  M},\\
	\end{align*}
	while $\mathcal{H}_i$ and $\mathcal{K}_i$ are given by
	\begin{align*}
		\mathcal{H}_1 =& -\frac{5 (16+9 \pi ) p (1323 \log 3-1276)}{373248 \pi  M^2}-\frac{(9 \pi -32) q (5 \log 3-4)}{486 \pi  M^2}\\
		& +\frac{4\left(200-117π+2187M²n₁\right)}{177147 \pi  M^4},\\
		\mathcal{K}_1 =& \frac{8\left(386-195π\right)}{177147 \pi  M^4}+\frac{5 p (4 (7756+4785 \pi )-405 (76+49 \pi ) \log 3)}{186624 \pi  M^2}\\
		& -\frac{q (916-1225 \log 3+60 \pi  (5 \log 3-4))}{972 \pi  M^2},\\
		\mathcal{H}_2 =& -\frac{4 (68+13 \pi )}{19683 \pi  M^4}+\frac{p (6380-6615 \log 3)}{41472 M^2}-\frac{(8+\pi ) q (5 \log 3-4)}{54 \pi  M^2},\\
		\mathcal{K}_2 =& \frac{2 (13 \pi -408)}{59049 \pi  M^4}+\frac{5 p (1323 \log 3-1276)}{248832 M^2}+\frac{(\pi -48) q (5 \log 3-4)}{324 \pi  M^2},\\
		\mathcal{H}_3 =& 0,\\
		\mathcal{K}_3 =& -\frac{2 (816+65 \pi )}{59049 \pi  M^4}+\frac{25 p (1276-1323 \log 3)}{248832 	M^2}-\frac{(96+5 \pi ) q (5 \log 3-4)}{324 \pi  M^2},\\
	\end{align*}
	Evaluated numerically, $\mathcal{X}_i,\mathcal{Y}_i,\mathcal{H}_i$ and $\mathcal{K}_i$ are given by
	\begin{align*}
		\mathcal{X}_1 =& \frac{1}{M}\left(0.0354 n₁ +0.0345 p+0.163 q\right)+\frac{0.00614}{M³},\\
		\mathcal{Y}_1 =& \frac{1}{M}\left(0.132p + 0.594q\right)+\frac{0.0336}{M³},\\
		\mathcal{X}_2 =& \frac{1}{M}\left(-0.124n₁+0.986p+0.188q\right)-\frac{0.00754}{M³},\\
		\mathcal{Y}_2 =& \frac{1}{M}\left(0.282p-0.0668q\right)-\frac{0.00285}{M³},\\
		\mathcal{X}_3 =& \frac{1}{M}\left(-0.124 n₁+0.0141p+0.0573q\right)-\frac{0.00831}{M³},\\
		\mathcal{Y}_3 =& \frac{1}{M}\left(0.113p+0.0635q\right)-\frac{0.00207}{M³},\\
		\mathcal{H}_1 =& -\frac{1}{M^2}\left(0.0157+0.0335 p+0.00364 q\right)-\frac{0.00120}{M^4},\\
		\mathcal{K}_1 =& -\frac{1}{M²}\left(-0.0951p+0.0486q\right)-\frac{0.00326}{M⁴},\\
		\mathcal{H}_2 =& -\frac{1}{M²}\left(0.0214p+0.0980q\right)-\frac{0.00704}{M⁴},\\
		\mathcal{K}_2 =& \frac{1}{M²}\left(0.00356p-0.0658q\right)-\frac{0.00396}{M⁴},\\
		\mathcal{H}_3 =& 0,\\
		\mathcal{K}_3 =& -\frac{1}{M²}\left(0.0178p+0.164q\right)-\frac{0.0110}{M⁴},\\
	\end{align*}
	Since the shell is made of branes, a massless gas (of open strings) and stiff matter made up of D0 branes, the stress energy tensor should be of the form  $S_{\textrm{total}}=S_{\textrm{brane}}+S_{\textrm{gas}}+S_{\textrm{stiff}}$ which determines $p$ and $q$.
	Up to order $k^{-1/2}$ these can be written as
	\begin{align*}
		p =& -\frac{32768 (5 \log 3-4) (5064752+9 \log 3 (622611 \log 3-1175608))}{3645 \sqrt{k} M^3 (371984+9 \log 3 (46593 \log 3-88552))^2}\\
		& -\frac{26624 (99 \log 3-92)}{1215 M^2 (371984+9 \log 3 (46593 \log 3-88552))}\\
		q =& \frac{128 (1323 \log 3-1276) (5064752+9 \log 3 (622611 \log 3-1175608))}{2187 \sqrt{k} M^3 (371984+9 \log 3 (46593 \log 3-88552))^2}\\
		& -\frac{832 (801 \log 3-884)}{81 M^2 (371984+9 \log 3 (46593 \log
			(3)-88552))},\\
	\end{align*}
	which evaluate to
	\begin{align*}
		p	&= -\frac{0.144}{M²}-\frac{0.423}{√kM³},\\
		q &= \frac{0.0162}{M²}-\frac{0.328}{√kM³}.
	\end{align*}
	The velocity vector $u^i$ at order $a²$ can be computed like in the first order case in \eqref{eq:ufirstorder} and is given by
	\begin{equation*}
		u^i = \left(3+γ a²,0,aβ\right),
	\end{equation*}
	where
	\begin{align*}
		γ =& -\frac{512}{2187 \sqrt{k} M^3}+\frac{16 (32+13 \pi ) \cos 2ϑ}{2187 \sqrt{k} M^3}+\frac{5 \pi  p (1323 \log 3-1276) \cos 2ϑ}{1152 \sqrt{k} M}\\
		& +\frac{2 \pi  q (5 \log 3-4) \cos 2ϑ}{3 \sqrt{k}
			M} -\frac{976 \cos 2ϑ}{729 M^2}-\frac{608}{729 M^2}-12 n₁+\frac{1}{16} q (44+25 \log 3)	\\
		& +\frac{5}{384} p (100+243 \log 3)\sin ²ϑ+\frac{1}{16} q (724-985 \log 3) \cos 2ϑ,
	\end{align*}
	which when evaluated numerically is
	\begin{align*}
		γ=&\frac{0.533 \cos 2ϑ}{\sqrt{k} M^3}-\frac{0.234}{\sqrt{k} M^3}+\frac{2.420 p \cos 2ϑ}{\sqrt{k} M}+\frac{3.13 q \cos 2ϑ}{\sqrt{k} M}-\frac{1.34 \cos 2ϑ}{M^2}-\frac{0.834}{M^2}\\
		& -12 n₁+4.78 p \sin ² ϑ-22.4 q \cos 2ϑ+4.47 q,
	\end{align*}
	and $β$ is given in \eqref{eq:ufirstorder}.
	This enables us to split the stress energy tensor into components and write \eqref{eq:stressenergysecondorder} as \eqref{eq:stressenergysplit} where $\mathcal{Z}_i$ provide $\mathcal{O}\left(a²\right)$ corrections to the first order results. The expressions are quite ugly, so instead of presenting the full analytical expressions, we present only the numerical values below
	\begin{align*}
		\mathcal{Z}₁ =& -\frac{1}{√kM⁴}\left(0.0276+0.00786M²n₁+0.0561\cos 2ϑ\right)\\
		& +\frac{1}{M³}\left(0.00812+0.0707M²n₁+0.000354\cos 2ϑ\right),\\
		\mathcal{Z}₂ =& -\frac{1}{M³}\left(0.0318+0.159M²n₁+0.0229\cos 2ϑ\right),\\
		\mathcal{Z}₃ =& -\frac{1}{√kM⁴}\left(0.0269+0.0157M²n₁+0.0191\cos 2ϑ\right).
	\end{align*}
	
	\small
	\bibliography{references}
	\bibliographystyle{utphys}
\end{document}